\documentclass[conference]{IEEEtran}
\IEEEoverridecommandlockouts
\usepackage{cite}
\usepackage{amsmath,amssymb,amsfonts}
\usepackage{algorithmic}
\usepackage{graphicx}
\usepackage{textcomp}
\usepackage{xcolor}
\def\BibTeX{{\rm B\kern-.05em{\sc i\kern-.025em b}\kern-.08em
    T\kern-.1667em\lower.7ex\hbox{E}\kern-.125emX}}

\usepackage{orcidlink}
\usepackage[nolist]{acronym}
\usepackage{pgfplots}
\usepgfplotslibrary{groupplots}
\pgfplotsset{compat=1.18}
\usepackage[font=footnotesize]{subcaption}
\usepackage[font=footnotesize]{caption}
\pgfplotsset{every axis/.append style={
	scaled x ticks = false,
	label style={font=\footnotesize},
	tick label style={font=\footnotesize},
	tick scale binop=\times
}
}
\usepackage{tabularx,environ}
\makeatletter
\newcommand{\problemtitle}[1]{\gdef\@problemtitle{#1}}% Store problem title
\newcommand{\probleminput}[1]{\gdef\@probleminput{#1}}% Store problem input
\newcommand{\problemoutput}[1]{\gdef\@problemoutput{#1}}% Store problem question
\NewEnviron{problem}{
  \problemtitle{}\probleminput{}\problemoutput{}% Default input is empty
  \BODY% Parse input
  \par\addvspace{.5\baselineskip}
  \noindent
  \begin{tabularx}{0.5\textwidth}{@{\hspace{\parindent}} l X c}
    \multicolumn{2}{@{\hspace{\parindent}}l}{\@problemtitle} \\% Title
    \textbf{Input:} & \@probleminput \\% Input
    \textbf{Output:} & \@problemoutput% Question
  \end{tabularx}
  \par\addvspace{.5\baselineskip}
}
\makeatother

\begin{document}

\title{Data Augmentation and Attention for massive MIMO-based Indoor Localization in Changing Environments\\
\thanks{The present work has received funding from the Smart Networks and Services Joint Undertaking (SNS JU) under the European Union's Horizon Europe Research and Innovation programme under Grand Agreement No 101192750 (6G-DALI).}
}

\author{\IEEEauthorblockN{Luisa Schuhmacher$^*$ \orcidlink{0000-0003-3293-7933},  Hazem Sallouha$^*$ \orcidlink{0000-0002-1288-1023}, Ihsane Gryech$^\dagger$ \orcidlink{0000-0001-5288-4205}, and Sofie Pollin$^{*,\mathsection}$ \orcidlink{0000-0002-1470-2076}}
\IEEEauthorblockA{$^*$ Department of Electrical Engineering (ESAT), KU Leuven, Leuven, Belgium,\\ $^\dagger$ Computer Science Department, Université Libre de Bruxelles (ULB), Brussels, Belgium, $^\mathsection$ Imec, Leuven, Belgium}
}

\maketitle
\begin{acronym}

\acro{ADN}{AttentionDenseNet}
\acro{AWGN}{additive white Gaussian noise}
\acro{DL}{Deep Learning}
\acro{DN}{DenseNet}
\acro{DNN}{Deep Neural Network}
\acro{CSI}{Channel State Information}
\acro{LoS}{Line-of-Sight}
\acro{MIMO}{Multiple-Input and Multiple-Output}
\acro{MSE}{Mean Squared Error}
\acro{nLoS}{non-Line-of-Sight}
\acro{SNR}{Signal-to-Noise Ratio}
\acro{ULA}{Uniform Linear Array}

\end{acronym}

\begin{abstract}
% Up to 250 words
The demand for high-precision indoor localization has grown significantly with the rise of smart environments, industrial automation, and location-aware applications. While massive Multiple-Input and Multiple-Output (MIMO) systems enable millimeter-level accuracy by leveraging rich Channel State Information (CSI), most existing solutions are optimized for static environments, where users or devices remain fixed during data collection and inference. Real-world applications, however, often require real-time localization in changing environments, where rapid movement, unpredictable blockages, and dynamic channel conditions pose significant challenges. To address these challenges, we introduce two data augmentation techniques designed to resemble blocked antennas, enhancing the generalizability of localization models to dynamic scenarios. Additionally, we enhance an existing Deep Learning (DL) model by incorporating attention modules, improving its ability to focus on relevant channel features and antennas. We train our model on data from a static scenario, augmented with the proposed techniques, and evaluate it on a dataset collected in changing scenarios. We investigate the performance enhancements achieved by the data augmentation techniques and the Attention modules, and observe a localization accuracy improvement from a mean error of 286 mm, when trained without Attention and without data augmentations, to 66 mm, when trained with Attention and data augmentation. This shows that high localization accuracy can be maintained in changing environments, even without training data from those scenarios.

\end{abstract}

\begin{IEEEkeywords}
Indoor Localization, Deep Neural Networks, Data Augmentation, Attention, Generalizability
\end{IEEEkeywords}

\section{Introduction}\label{introduction}
The demand for high-precision indoor localization has surged with the rise of smart environments, industrial automation, and location-aware applications \cite{hailu_indoor_2025, kerdjidj_uncovering_2024}. Massive \ac{MIMO} systems, with their high angular resolution and rich \ac{CSI}, have emerged as a key enabler for achieving millimeter-level accuracy in indoor settings \cite{tian_high-precision_2023}. While traditional time- and angle-based localization techniques struggle in complex real-world conditions, entailing \ac{nLoS} or low \ac{SNR} scenarios, data-driven approaches have demonstrated remarkable success \cite{tang_regression_2019, dai_deepaoanet_2021}. These methods leverage either raw \ac{CSI} \cite{de_bast_expert-knowledge-based_2022, li_toward_2022} or features generated from \ac{CSI} based on expert knowledge \cite{dai_deepaoanet_2021, tian_deep-learning-based_2024, tian_attention-aided_2024} to extract meaningful patterns, often outperforming classical geometric-based techniques \cite{li_toward_2022, xu_swin-loc_2024}.

Despite these advancements, most data-driven localization methods are still tailored to static environments, where users or devices remain stationary during data collection and inference \cite{hejazi_dyloc_2021}. However, real-world applications such as tracking mobile robots, wearable devices, or users in dynamic spaces require real-time localization in changing environments. These scenarios introduce challenges like rapid movement, unpredictable blockages, and continuously changing channel conditions, which can significantly degrade the performance of localization models trained on data from static environments.

To address localization in dynamic environments, frameworks like DyLoc \cite{hejazi_dyloc_2021} have been proposed, transforming \ac{CSI} into angle-delay profiles and employing recurrent neural networks to track user movement. However, DyLoc and its extensions \cite{nguyen_deep_2023, nguyen_efficient_2023} still rely on simulated dynamic data derived from static measurements. They assume the availability of initial calibration samples, which may not be practical in truly changing environments. Additionally, while data augmentation techniques such as noise injection and synthetic multipath generation have improved model generalization \cite{dai_deepaoanet_2021}, they have not been specifically tailored to simulate \ac{nLoS}, a common issue in changing indoor settings.

In this paper, we introduce a data-driven localization solution for dynamic indoor environments. In particular, we propose two novel data augmentation techniques to simulate blocked antennas, enhancing the robustness of localization models in changing environments. We further go beyond the current state-of-the-art by inserting Attention \cite{vaswani2017attention} modules in an existing \ac{DL} model \cite{bast_csi-based_2019}, thereby improving the model's ability to focus on relevant channel features as well as antennas under \ac{LoS} conditions. To assess the performance of our proposed localization approach, we trained our model on a static environment, augmented with our proposed techniques, and evaluated it on data collected in a changing environment. Our key contributions are as follows:

\begin{itemize}
    \item Data augmentation for \ac{nLoS} resulting from blocked antennas: We propose two data augmentation methods to resemble blocked antennas, and compare their influence on the model's robustness to dynamic scenarios.
    \item Attention-enhanced \ac{DL} model: We enhance an existing state-of-the-art \ac{DL} model by incorporating Attention modules, enabling it to better adapt to changing and dynamic channel conditions.
    \item Evaluation in changing environments: We evaluate our model on a dataset containing changes in the environment, demonstrating its effectiveness in maintaining high localization accuracy despite not having seen any data collected from an environment containing blockages during training.
\end{itemize}

Code accompanying the paper is made open-source\footnote{https://gitlab.kuleuven.be/networked-systems/public/data-augmentation-and-attention-for-csi-based-indoor-localization-in-changing-environments}. The remainder of this paper is organized as follows: Section \ref{related_work} reviews related work on indoor localization and \ac{CSI} data augmentation. Section \ref{methodology} details the system setup, our proposed data augmentation techniques and model enhancements. Following, Section \ref{performance_evaluation} presents experimental results and analyzes the impact of the data augmentation techniques and the Attention modules. Finally, Section \ref{conclusion} concludes the paper with a discussion and future directions.

\section{Related Work}\label{related_work}
Recent indoor localization advances have leveraged classical signal processing and \ac{DL} techniques. Classical angle-based methods such as MUSIC and ESPRIT are limited by their reliance on \ac{LoS} channel conditions, often failing in \ac{nLoS} or low \ac{SNR} scenarios \cite{dai_deepaoanet_2021}. In contrast, \ac{DL}-based approaches have demonstrated superior performance by directly learning from raw \ac{CSI} or covariance matrices, achieving centimeter- or even millimeter-level accuracy in controlled environments \cite{bast_csi-based_2019, de_bast_expert-knowledge-based_2022, li_toward_2022, xu_swin-loc_2024}. Specifically, Attention mechanisms have further improved localization accuracy by their ability to focus on relevant channel features. Swin-Transformers achieved an \ac{MSE} as low as 8.46 mm for \ac{ULA} setups \cite{xu_swin-loc_2024}. In the case of distributed antenna systems, attention-based convolutional networks reduced the mean error to 5.78 mm \cite{ma_millimeter_2024}.

A recent trend to enhance the generalizability of \ac{DL} models is to enhance training data with data augmentations. Techniques such as adding artificial noise, phase shifts, and synthetic multipath components have been used to simulate diverse channel conditions \cite{dai_deepaoanet_2021, ma_millimeter_2024}. For instance, the authors in \cite{dai_deepaoanet_2021} augmented their dataset with \ac{AWGN} and phase shifts to emulate varying \ac{SNR}s and multipath effects, resulting in models that generalize well to unseen environments. Their goal, however, is to predict the Angle-of-Arrival, rather than the user's position, resulting in a different problem formulation. Feature construction from multiple domains (e.g., Doppler, delay, and spatial) has also been explored. The authors in \cite{mukherjee_multi-domain_2025} demonstrated that incorporating Doppler domain features improves prediction accuracy.

Although data-driven localization in static environments has seen significant progress, indoor localization in dynamic environments, where users or objects move rapidly, remains underexplored. To the best of our knowledge, \cite{hejazi_dyloc_2021} is the only work that proposes a data-driven method for \ac{CSI}-based indoor localization in dynamic scenarios. They transform \ac{CSI} into angle-delay profiles and use recurrent neural networks to track user movement. However, they view the problem as tracking, therefore always having a time-series of \ac{CSI} snapshots available for localization. We, in contrast, aim to localize from \ac{CSI} from a single point in time, using a model trained on data from a previous, but static environment.

Additionally, one study examined the performance of their \ac{DNN} when evaluated on data from a changing scenario, despite having been trained only on data from a static scenario \cite{bast_mamimo_2020}. To achieve this, an open dataset was utilized that contains both a dense dataset collected in a static environment using \ac{CSI} from a multitude of different user positions and a nomadic dataset. The nomadic dataset was collected in the same room with the same antenna setup across four different user positions. During data collection, humans walked along specific trajectories, obstructing \ac{LoS} at some instances and introducing signal reflections \cite{dataset} at others. A performance degradation of 2 to 5 times from their initial millimeter-level positioning accuracy was reported for movements that are not blocking \ac{LoS}, and a substantial performance degradation of up to 50 times in scenarios where the human is blocking \ac{LoS}.

In this work, we tackle the issue of the inability of indoor localization \ac{DL} models to generalize from static to dynamic environments \cite{bast_mamimo_2020}. We therefore propose two data augmentations, along with a new model featuring Attention, to strengthen performance in unseen changing scenarios. To the best of our knowledge, no work so far has systematically addressed instantaneous \ac{CSI}-based localization without prior expertise in changing environments, where \ac{LoS} and \ac{nLoS} channel characteristics change unpredictably due to user mobility, blockages, and varying \ac{SNR} conditions. This gap motivates our work, which aims to develop robust, adaptive models that are trained in static environments yet capable of generalizing to dynamic ones.

\section{Data Augmentation-aided Attention-based Indoor Localization}\label{methodology}
We introduce two innovations to generalize \ac{CSI}-based indoor localization to changes in the environment. First, we propose to enhance the collected \ac{LoS} data from a static environment with augmented data, designed to mimic blocked antennas. To this end, we implement a vanilla approach of setting \ac{CSI} collected at random antennas to zero. This, however, discards all information, which is not the case in \ac{nLoS} scenarios. We therefore propose an alternative data augmentation approach that randomly attenuates the received \ac{CSI}.

Second, we adopt the architecture introduced in \cite{bast_csi-based_2019}, which has been shown to achieve millimeter-level localization performance in static scenarios. We enhance this architecture by inserting two Attention modules, enabling the model to learn which antennas and subcarriers contain more helpful information, and leveraging this understanding to localize the user more accurately.

\subsection{System Setup}
In our setup, we consider an indoor massive \ac{MIMO} system where a 64-antenna base station estimates \ac{CSI} from pilot signals transmitted by a single-antenna user. The \ac{CSI} is represented as a complex matrix $H_\text{CSI} \in \mathbb{C}^{64 \times 100}$, with 64 antennas deployed as a \ac{ULA}, receiving signals over 100 evenly spaced subcarriers. Users are precisely positioned with a positional error of less than 1 mm at four distinct positions, with a minimum distance of 1.5 m between any two users. The system operates at a center frequency of 2.61 GHz, giving the wavelength $\lambda=114.56$ mm. It utilizes a bandwidth of 20 MHz, resulting in a single-antenna range resolution of approximately 7.5 m. The \ac{ULA} antenna elements are spaced 70 mm apart, resulting in a $0.61 \lambda$-spacing. The base station antennas are mounted 93 cm above the floor, while the user antennas are placed 20 cm above the floor.

\subsection{Data Augmentations}
In the following, we outline two data augmentations to enhance the model's generalizability.

\subsubsection{Vanilla}
The goal is to introduce blocked components into a data set that contains only data collected in a \ac{LoS} setting. A simple way to do that is to completely zero out a subset of antennas, thereby deleting all information those antennas contain. The subsequent model should then learn to identify antennas that contain useless data, discard those and only use antennas in \ac{LoS} to localize the user. For this vanilla approach, we randomly choose half of the training samples. We select a random subset of arbitrary size from the antennas for these samples, set the \ac{CSI} at these antennas to zero, and add the augmented sample to the training dataset. Note that while direct \ac{LoS} blockage by a human initially suggests that neighboring antennas are affected, other factors, such as reflections, contradict this assumption, supporting our random choice.

\subsubsection{Random Attenuation}
As loss of all information does not accurately reflect the \ac{nLoS} resulting from the dynamic environment, we propose random attenuation as a second data augmentation method. Similar to the vanilla data enhancements, we select a random subset of arbitrary size from the antennas of half of the training samples. We then scale the received signal by a random factor, which attenuates the signal between 10 dB and 40 dB. While the lower range, 10-20 dB, is typically observed as an attenuation of human blockage \cite{Cui2019Influence}, we extend this range to 40 dB to account for rare edge cases. The augmented samples are again added to the training data. The model is then expected to learn to extract useful information from both \ac{LoS} \ac{CSI}, which is all the data that has not been augmented, and attenuated \ac{CSI}, the augmented values.

\subsection{Model}
\begin{figure}
    \centering
    \includegraphics[width=0.8\linewidth]{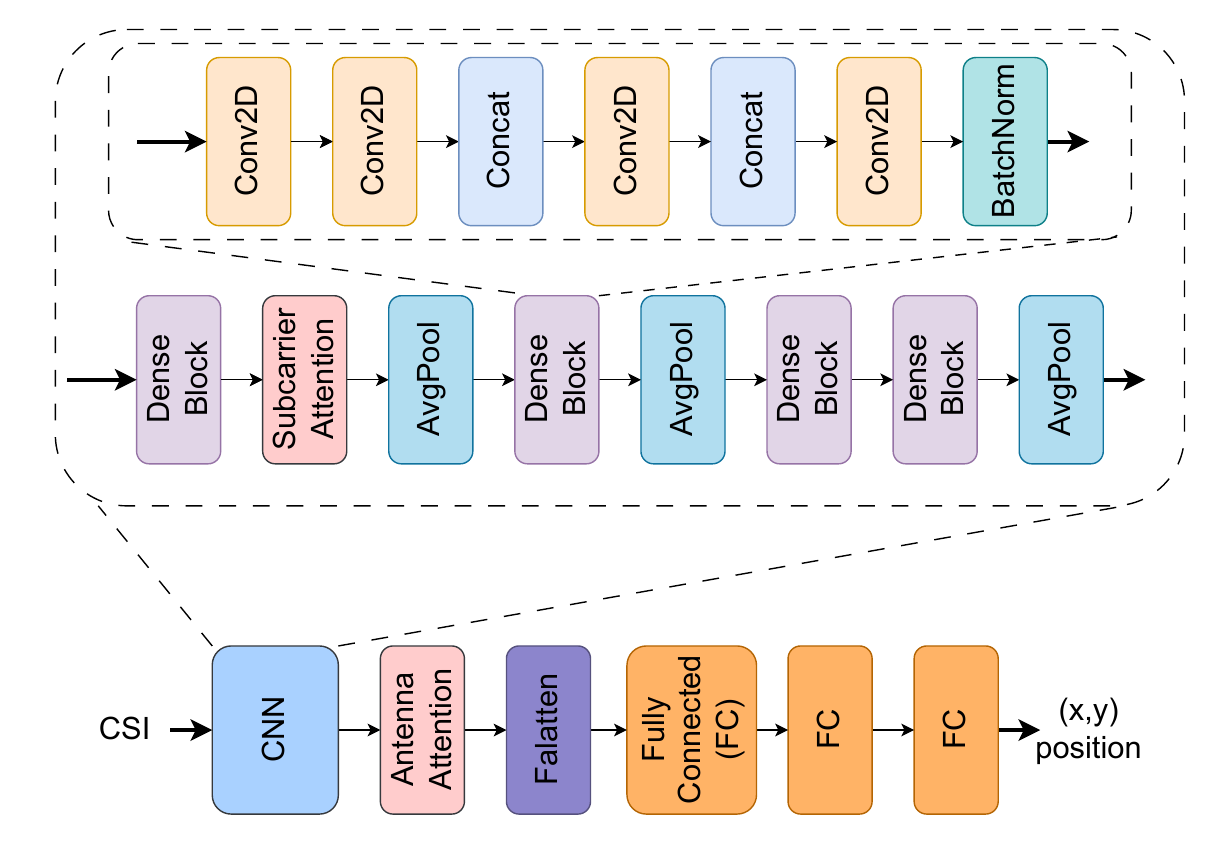}
    \caption{Our proposed enhancement of the \ac{DNN} in \cite{bast_csi-based_2019}. We insert two Attention modules, shown in pink, to enable the model to focus on specific antennas and subcarriers. The rest follows the original model architecture.}
    \label{fig:models}
\end{figure}

In addition to data augmentations, we propose an enhancement to the \ac{DNN} deployed in \cite{bast_csi-based_2019}. Specifically, we introduce two Attention modules \cite{vaswani2017attention}. Attention is a well-known mechanism first introduced in Natural Language Processing, which allows modern Large Language Models to assist us with a variety of tasks. In a nutshell, Attention, or Self-Attention to be precise in terminology, is given by the following formula:
\begin{equation*}
    \text{Attention}(Q,K,V)=\text{softmax}\left(\frac{QK^T}{\sqrt{d_k}}\right)V,
\end{equation*}
where $Q$, $K$ and $V$ are matrices called Query, Key and Value, and $d_k$ notes the dimension of $K$. The three matrices get produced by multiplying the embedding $E$ generated by the previous layers of the \ac{DNN} to learnable weight matrices $W_Q$, $W_K$ and $W_V$, which is expressed as
\begin{equation*}
    Q=W_QE,\ K=W_KE,\ V=W_VE.
\end{equation*} 

The first Attention module we introduce is designed to put Attention on the subcarriers. Due to subcarrier-specific phenomena such as frequency-selective fading, the \ac{CSI} at some subcarriers may contain more information about the user position than others. Therefore, an Attention block is executed for each subcarrier after the model has learned a meaningful embedding for the \ac{CSI} on a subcarrier level. Using the resulting new embedding, the initial architecture extracts embeddings on the antenna level. After that, we insert the second Attention module, which puts Attention on a higher level, specifically the antennas. Using this, the model can learn to neglect (partially) blocked antennas, and mainly use the \ac{CSI} from antennas with \ac{LoS}. Lastly, the initial model architecture extracts the location from the learned embeddings. A visualization of the model with the Attention modules can be seen in Figure \ref{fig:models}. The pink blocks represent the newly inserted Attention modules; the rest of the architecture is the original one deployed in \cite{bast_csi-based_2019}. We will refer to this enhanced version of the model in \cite{bast_csi-based_2019} in the following as \textit{\ac{ADN}}.

\section{Performance Evaluation}\label{performance_evaluation}
\begin{figure}
    \centering
    \includegraphics[width=0.78\linewidth]{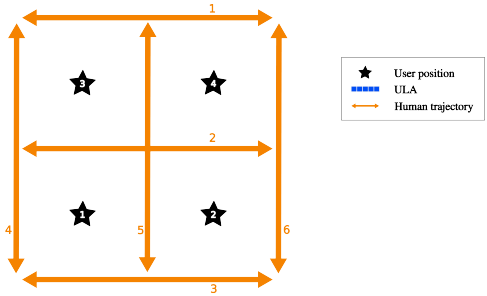}
    \caption{The setting of the nomadic dataset \cite{dataset}. The black stars show the user positions, in blue the \ac{ULA}, and in orange six trajectories of a human walking, each defining a new scenario in which data is collected. For the static scenario, no human is moving.}
    \label{fig:data}
\end{figure}

To evaluate our proposed data augmentations and model modifications, we utilize the nomadic dataset from a publicly available dataset \cite{dataset}. This is the same dataset that has been leveraged by the study on performance degradation of \ac{DL} models trained on a static scenario when evaluated on changing scenarios \cite{bast_mamimo_2020}. We use the sub-dataset, which contains \ac{CSI} collected at a \ac{ULA} with 64 antennas over 100 subcarriers, each from four different user positions and in 7 different scenarios. One scenario is static, meaning no changes in the environment occur at the time of data collection. The other six scenarios are changing and include a human walking a specific trajectory up and down during data collection. A depiction of the setting is shown in Figure \ref{fig:data}. Note that, although we have only four different user positions, for generalizability, we model the problem as a regression problem rather than a classification problem.

To evaluate the impact of the data augmentations and the additional improvements expected from our proposed model, \ac{ADN}, we train and test the initial \ac{DNN} in \cite{bast_csi-based_2019} as a baseline. We refer to it as \textit{\ac{DN}} in the following. For both model training of the \ac{DN} and the \ac{ADN}, we split the static scenario into a train-val-test split of 70-15-15, and, if applicable, add the proposed data augmentation to the training data. We train \ac{DN} and \ac{ADN} ten times each with different random initializations and select the model that achieves the highest static test accuracy for evaluation on the changing scenarios.

\subsection{Static Performance}
\begin{table}[]
    \caption{Mean error over the static testset. For each data augmentation method, the \ac{DN} (DN) and \ac{ADN} (ADN) architectures were trained 10 times. The models that achieved the lowest test error are reported here. RA abbreviates Random Attenuation.}
    \centering
    \begin{tabular}{|l|c|c|c|c|c|c|}
        \hline
        \textbf{Data augm.} & \multicolumn{2}{c|}{None} & \multicolumn{2}{|c|}{Vanilla} & \multicolumn{2}{c|}{RA} \\
        \hline
        \textbf{Model} & DN & ADN & DN & ADN & DN & ADN \\
        \hline
        \textbf{Test error} & 6 mm & 8 mm & 15 mm & 18 mm & 4 mm & 8 mm \\
        \hline
    \end{tabular}
    \label{tab:static_perf}
\end{table}
We begin by presenting the performance of models trained with and without data augmentations in the static scenario. Table \ref{tab:static_perf} shows the mean localization error on the testset for the two models trained without data augmentations, with the vanilla data augmentation, and with random attenuation. Note that no data augmentations are present in the test set. 

Surprisingly, in the static scenario, the \ac{DN} consequently outperforms the \ac{ADN}. This can be attributed to the fact that the initial architecture already yields a very low mean localization error of 5-15 mm, indicating that it is well-designed for the task. Therefore, adding more blocks to the architecture can lead to a performance drop \cite{frankle2019lottery}. We further observe that training with the vanilla data augmentations worsens performance in the static case, showing that they do not help generalize to unseen \ac{CSI} collected in an unchanged environment. When introducing random attenuation, however, the performance of the \ac{ADN} remains unchanged, whereas the performance of \ac{DN} improves. This suggests that random attenuation helps the model generalize better within the same environment.

\subsection{Performance in changing Environments}
\begin{figure}
    \centering
    \includegraphics[width=0.75\linewidth]{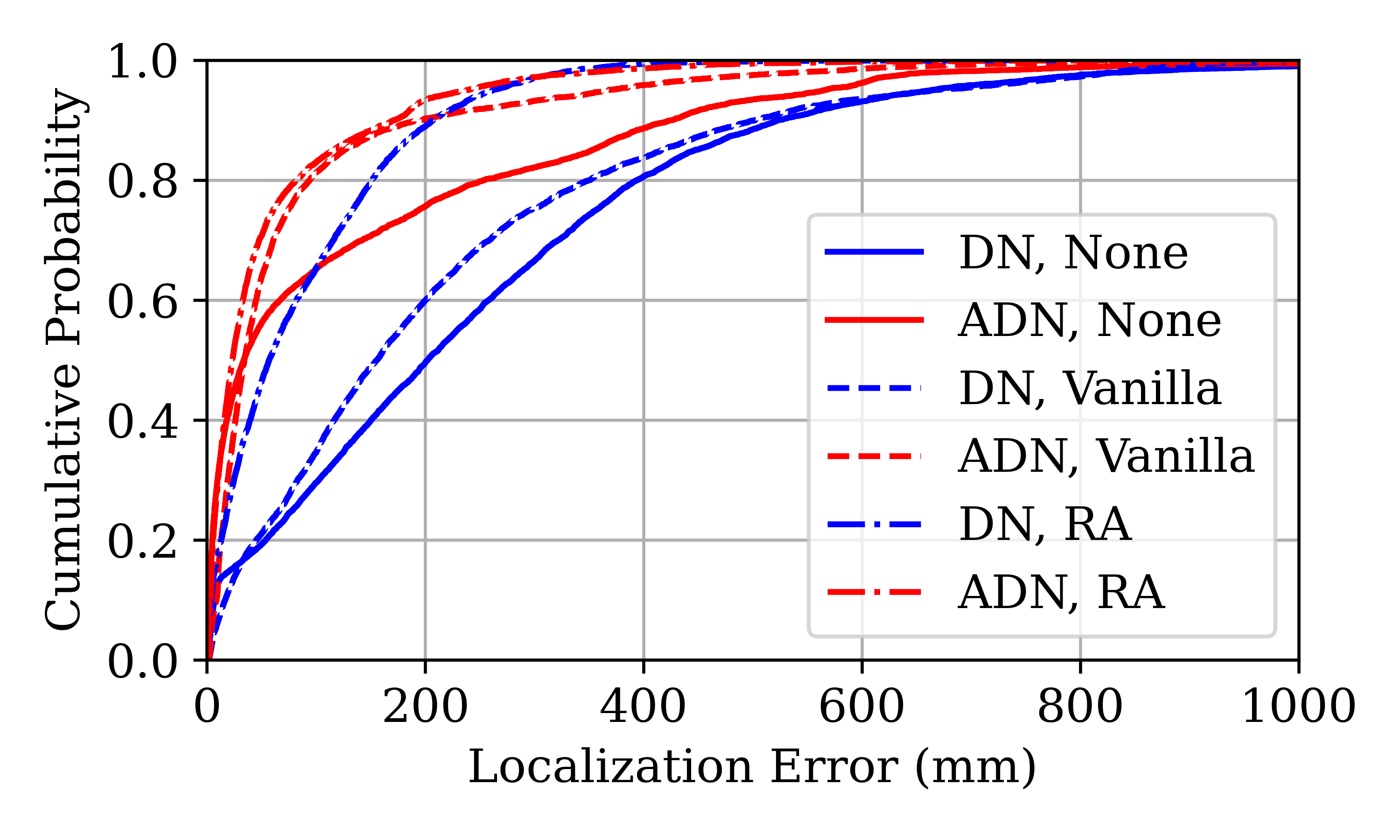}
    \caption{Cumulative distribution function of \ac{DN}'s and \ac{ADN}'s localization error, both trained either without data augmentations (None), with the vanilla data augmentations (vanilla) or with random attenuation (RA).}
    \label{fig:cdfs}
\end{figure}
\pgfplotsset{compat=1.11,
    /pgfplots/ybar legend/.style={
    /pgfplots/legend image code/.code={%
       \draw[##1,/tikz/.cd,yshift=-0.25em]
        (0cm,0cm) rectangle (3pt,0.8em);},
   },
}
\begin{figure}
    \centering
    \begin{tikzpicture}
\begin{groupplot}[
    group style={
        group size=2 by 1,
        horizontal sep=5pt,
        vertical sep=5pt,
        x descriptions at=edge bottom,
        y descriptions at=edge left,
    },
    legend style={font=\scriptsize},
    symbolic x coords={None, Vanilla, RA, Upper Bound},
    xtick=data,
    ylabel=Test error (mm),
    ymin=0,
    ymax=350,
    enlarge x limits=0.3,
    legend style={at={(0.34,0.94)},
    anchor=north,legend columns=-1},
    ybar,
    nodes near coords,
    nodes near coords align={vertical},
    nodes near coords style={font=\scriptsize},
    height=3.5cm,
]

\nextgroupplot[
    xmin=None, xmax=RA,
    xtick={None, Vanilla, RA},
    ylabel=Test error (mm),
    width=6cm,
    bar width=15pt
]
\addplot coordinates {(None,286) (Vanilla,248) (RA,100)};
\addplot coordinates {(None,156) (Vanilla,91) (RA,66)};

\nextgroupplot[
    xmin=Upper Bound, xmax=Upper Bound,
    xtick={Upper Bound},
    yticklabels={},
    ylabel={},
    width=3cm,
    bar width=15pt
]
\addplot coordinates {(Upper Bound,13)};
\addplot coordinates {(Upper Bound,12)};
\legend{\ac{DN}, \ac{ADN}}

\end{groupplot}
\end{tikzpicture}
    \caption{Mean error when evaluating the models on the changing scenarios. In blue, the \ac{DN}, and in red, the \ac{ADN}, both trained with either no data augmentations (None), vanilla data augmentations (Vanilla), random attenuation (RA), or directly on data from the changing scenarios (Upper Bound). The last case serves as an indication of how well the models can perform, and is only evaluated on a subset of the data, as the rest has been used for training.}
    \label{fig:nomadic_perf}
\end{figure}
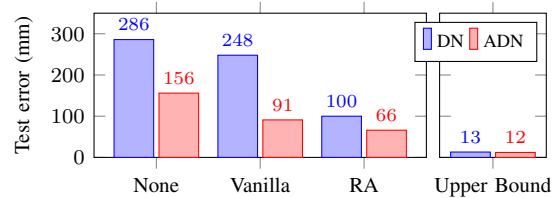
We now examine the models' performance in the changing scenarios. We therefore pick the model with the lowest test error in the static scenario for each model architecture and data augmentation combination. We emphasize that the models were not exposed to any data resembling the changing scenarios during training, so this evaluation particularly assesses their generalization capabilities. Figure \ref{fig:cdfs} and \ref{fig:nomadic_perf} depict the results in terms of the error's cumulative distribution function and mean error, respectively. In Figure \ref{fig:nomadic_perf}, we additionally show the performance of both \ac{DN} and \ac{ADN} when trained on data from the changing scenarios directly, establishing an upper bound of 13 mm for \ac{DN} and 12 mm for \ac{ADN}. These benchmarks highlight the potential of these architectures when exposed to the target scenarios during training.

Figure \ref{fig:cdfs} shows that for both data augmentation techniques, the \ac{ADN} consistently outperforms its \ac{DN} counterpart, demonstrating the added value of incorporating attention modules into the architecture for unseen scenarios. We further observe in Figure \ref{fig:nomadic_perf} that for both \ac{DN} and \ac{ADN}, training with the vanilla data augmentations yields a lower mean evaluation error than training without them. This shows that randomly zeroing out \ac{CSI} adds enough diversity for the model to generalize better in the changing scenarios. Finally, randomly attenuating antennas, and therefore not discarding all information that has been there before, shows the best performance. The \ac{ADN} trained with random attenuation data augmentations has a mean test error on the changing scenarios of 66 mm, further reducing the error of its counterpart trained on the vanilla data augmentations by 27\%. We also observe from Figure \ref{fig:nomadic_perf} that the \ac{DN} trained with random attenuation performs worse than the \ac{ADN} trained on the vanilla data augmentations. In contrast, it performs better than the \ac{ADN} trained without data augmentations. This showcases that well-designed data augmentations can be powerful for a model's generalizability in unseen scenarios, outperforming simply using an optimized architecture.

Having evaluated the overall performance, we continue with an investigation of the models' step-by-step performance over time for a given changing scenario.

\subsubsection{Performance analysis in a selected changing scenario}
\begin{figure}[]
  \centering
  \begin{subfigure}{0.24\textwidth}
    \centering
    \includegraphics[width=\textwidth]{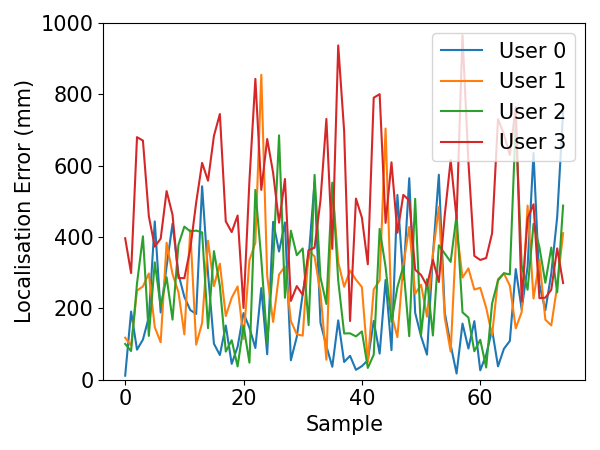}
    \caption{\ac{DN} trained without data augmentations.}
  \end{subfigure}
  \begin{subfigure}{0.24\textwidth}
    \centering
    \includegraphics[width=\textwidth]{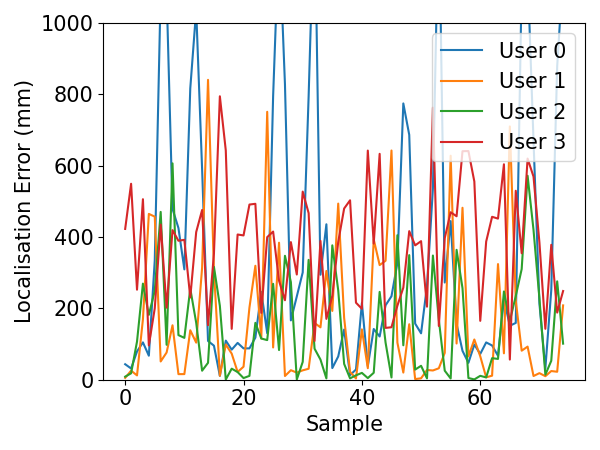}
    \caption{\ac{ADN} trained without data augmentations.}
  \end{subfigure}
  \begin{subfigure}{0.24\textwidth}
    \centering
    \includegraphics[width=\textwidth]{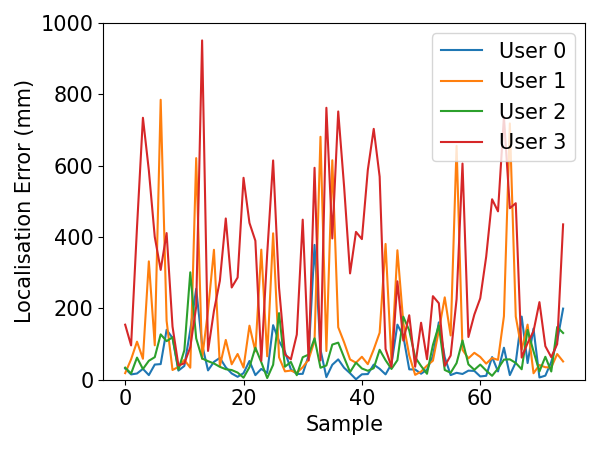}
    \caption{\ac{ADN} trained with vanilla data augmentations.}
  \end{subfigure}
  \begin{subfigure}{0.24\textwidth}
    \centering
    \includegraphics[width=\textwidth]{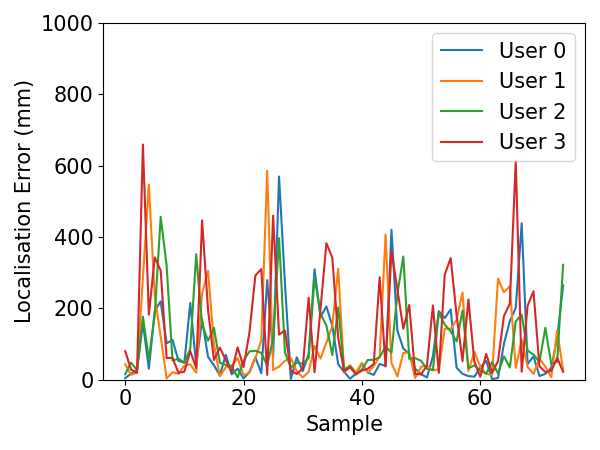}
    \caption{\ac{ADN} trained with random attenuation.}
  \end{subfigure}
  \caption{Example error curves of different models, trained with different data augmentation techniques, evaluated on the changing scenario where a human walks up and down between the \ac{ULA} and the four users. \ac{ADN} trained with random attenuation shows the most stable behavior among all four users (d).}
  \label{fig:error_curves}
\end{figure}
We now examine how the models perform for one of the changing scenarios. In the chosen scenario, a person walks back and forth between the \ac{ULA} and the users, occasionally blocking the \ac{LoS} path for specific antennas while creating reflections for others. Although the data does not include the ground-truth position of the moving person at each sample, we are aware of the general back-and-forth movement. Consequently, we can expect a certain periodicity in the error curves (i.e., when the human blocks \ac{LoS}, we expect a higher error than when the human is far away from the direct path between the user and the antenna, and therefore, limits its influence on the captured \ac{CSI}). The curves are shown in Figure \ref{fig:error_curves}. Due to space limitations, we omit the \ac{DN} trained with the vanilla data augmentation and the \ac{DN} trained with random attenuation, and only show the evolution of the error for the \ac{DN} trained without data augmentations (Figure \ref{fig:error_curves}a), which is the model with the worst overall performance, and our \ac{ADN} trained without data augmentations (Figure \ref{fig:error_curves}b), with the vanilla data augmentations (Figure \ref{fig:error_curves}c), and with random attenuation (Figure \ref{fig:error_curves}d). Following its best overall performance, \ac{ADN} trained with random attenuation (Figure \ref{fig:error_curves}d) demonstrates stable behavior between all four users, with compared to the other models, low base error and error peaks.

For all four error curves, we can see the expected periodic spikes. When comparing the performance of the \ac{DN} (Figure \ref{fig:error_curves}a) and the \ac{ADN} (Figure \ref{fig:error_curves}b) when trained without data augmentations, we observe an interesting pattern: \ac{ADN} exhibits a lower base error for user 1 and user 2, but higher peaks for user 0. This indicates that the attention modules, when trained without any data augmentations, manage to filter out minor disturbances, like those added from reflection, but might struggle with significant disturbances, \ac{nLoS}, and even worsen performance in those cases. 

We now, therefore, look at the error curves of the \ac{ADN}s trained with the vanilla data augmentations (Figure \ref{fig:error_curves}c) and with random attenuation (Figure \ref{fig:error_curves}d). It can be seen that the \ac{ADN} trained with the vanilla data augmentations maintains the localization error for users 0 and 2 below 200 mm, with one outlier in each case. In contrast, for users 1 and 3, the periodic peaks are significantly higher, reaching values between 400 mm and 1000 mm. There is no geometric reason in the data collection setup to explain this behavior. This, therefore, indicates randomness and suggests that generalizability from training with vanilla data augmentations is not necessarily given. Finally, the \ac{ADN} trained with random attenuation, which yielded the best mean error, also exhibits the most stable error curves. The periodic peaks are still visible, but barely exceed 500 mm, and therefore are lower than those from the other models (except for the two very well-localized users by the \ac{ADN} trained with the vanilla data augmentations). Furthermore, all four users show similar curves, indicating a steady and more trustworthy performance.

It is worth noting that the error peaks remain relatively high, reaching up to 500 mm, compared to the static mean test error of 4 mm (for the best model). Random attenuation does not capture all the complex influence a human walking in a room has on the captured \ac{CSI}. But it shows that with this simple data augmentation, localization accuracy from \ac{DNN}s can be drastically improved in unseen scenarios.

\subsubsection{Attention module analysis}
\begin{figure}
    \centering
    \begin{subfigure}{0.24\textwidth}
        \includegraphics[width=\textwidth]{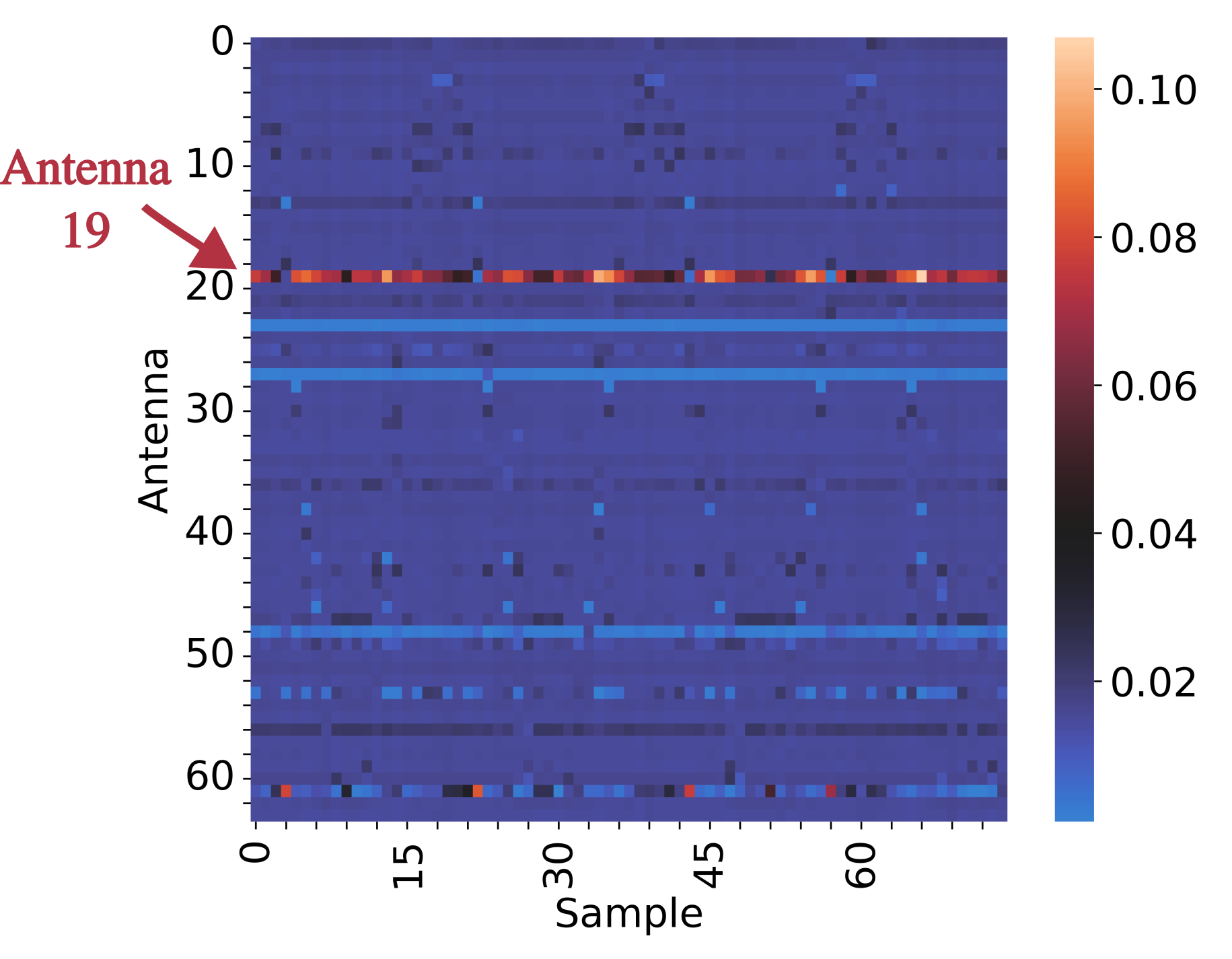}
    \end{subfigure}
    \begin{subfigure}{0.24\textwidth}
        \includegraphics[width=\textwidth]{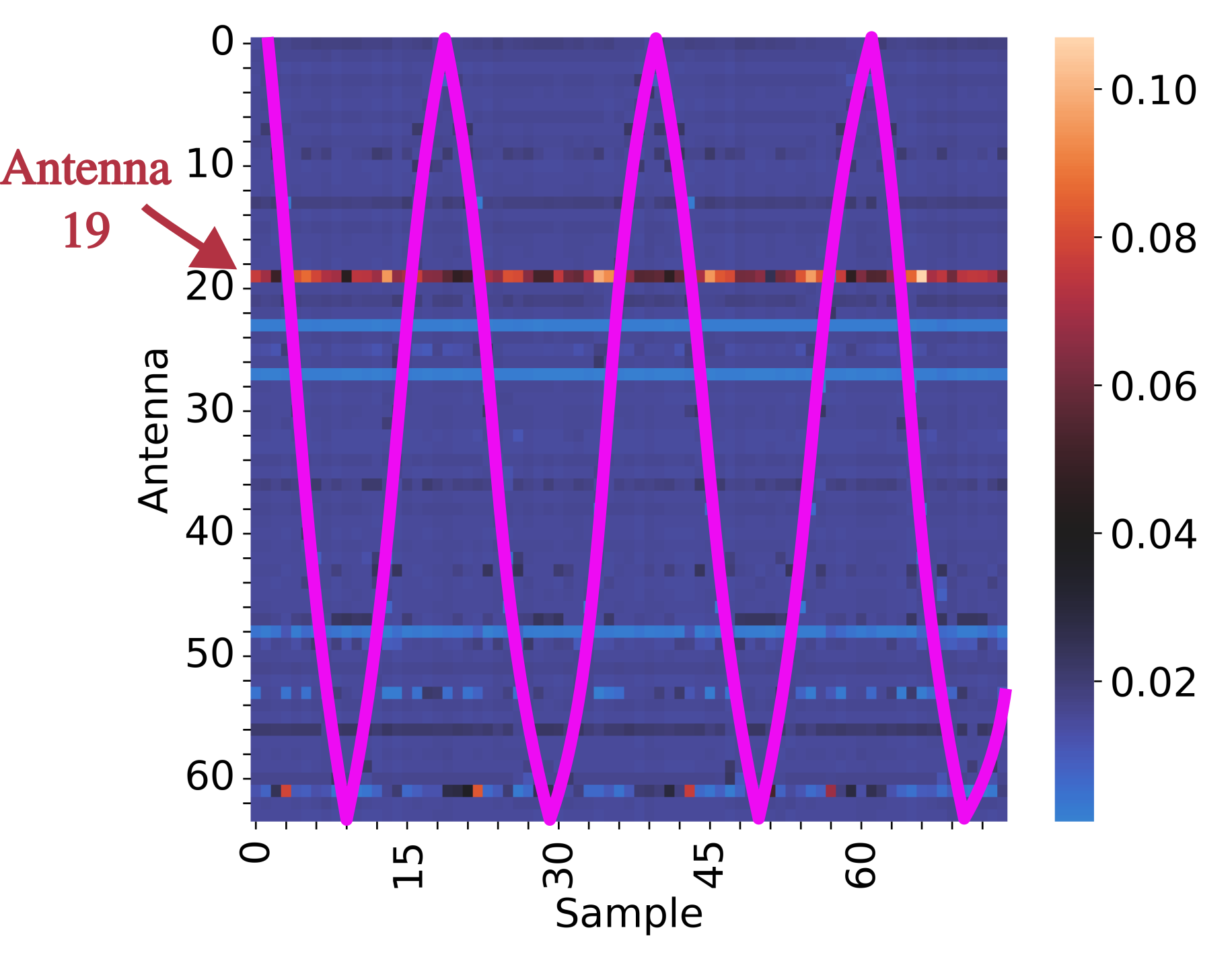}
    \end{subfigure}
    \caption{Visualization of the antenna Attention module of the \ac{ADN} trained with random attenuation. The heatmaps show the weight calculated by the Attention module for each antenna. On the x-axis, the samples over time for the scenario where the human is walking up and down between the \ac{ULA} and the users. On the y-axis are the 64 antennas. We depict the same heatmap twice: once in its original form (left) and once with the pattern of human movement emphasized (right). The Attention block assigns most weight to antenna 19, which is closest to the user location.}
    \label{fig:heatmap}
\end{figure}
Lastly, we investigate the behavior of the Attention module. For simplicity, we only look at the antenna Attention module. We chose the \ac{ADN} trained with random attenuation for this analysis, which showed the best performance. 

Figure \ref{fig:heatmap} shows a heatmap of the weights calculated by the antenna Attention module for each antenna and for each sample from the scenario where a human is walking back and forth between the \ac{ULA} (trajectory 3 in Figure \ref{fig:data}) and user 3. It can be seen that most of the weight goes to antenna 19. When examining the selected scenario (trajectory 3 and user 3 in Figure \ref{fig:data}), this aligns with the antenna closest to the user, which therefore receives the \ac{CSI} with the highest signal strength. 

For some samples, however, a lower weight is given to antenna 19. In fact, upon closer examination of the heatmap, a pattern emerges where the calculated weight for an antenna differs from the weights of the samples surrounding it. This pattern strongly resembles the expected human movement. It is worth noting that we do not have ground truth as to where the human is standing precisely for each sample; however, the pattern indicates which antennas are blocked in each sample and shows the human walking back and forth a couple of times. This shows that the Attention module captures both the antenna containing the most information and the human movement, and adjusts the given weights accordingly.

\section{Conclusion \& Future Work}\label{conclusion}
We introduced two data augmentation techniques: setting \ac{CSI} at random antennas to zero and attenuating \ac{CSI} at random antennas. We further adapted a \ac{DNN} for indoor localization from the literature, and integrated two Attention modules, operating on the subcarrier level and the antenna level, respectively. We trained the initial architecture and our proposed architecture on data collected in a static environment, but enhanced with either of the proposed data augmentations. Our results showed that solely the Attention modification, without data augmentation-enhanced training, enabled a higher localization accuracy in a changing environment, reducing the mean localization error by nearly half, from 286 mm to 156 mm. Training with data augmentations also improved generalizability for both \ac{DNN} architectures, whereas training with random attenuation gave the best results. The best-performing model is the \ac{DNN} architecture with Attention, trained on data augmented by introducing random attenuation, with a mean localization error of 66 mm. 

While our study has some limitations, most notably that the dataset includes only four user positions, it effectively demonstrates a proof of concept by framing the problem as a regression task rather than a classification task. We further note that human movement in our scenarios is restricted to fixed, simple trajectories, and that the effectiveness of random attenuation as data augmentation in a more varying environment remains to be studied. In the future, we plan to collect a more diverse dataset to evaluate our method for users with a broader range of positions. Further, when trained directly on the changing scenarios, the best model achieves a mean localization error of 12 mm, whereas in the changing environment, this error rises to 66 mm for the best model. This leaves room for further improvement in data augmentations to enable generalizability, which we also plan to tackle in future work.

\bibliographystyle{ieeetr}
\bibliography{references}

@article{frankle2019lottery,
    author = {Frankle, Jonathan and Carbin, Michael},
    title = {The Lottery Ticket Hypothesis: Finding sparse, trainable neural neworks},
    journal = {ICLR},
    year = {2019}
}

@article{Cui2019Influence,
title={Influence of Human Body on Massive {MIMO} Indoor Channels},
author={P.-F. Cui and others},
journal={IEEE VTC-Spring},
year={2019}
}

@article{hailu_indoor_2025,
AUTHOR = {Hailu, Tesfay Gidey and Guo, Xiansheng and Si, Haonan},
TITLE = {Indoor Positioning Systems as Critical Infrastructure: An Assessment for Enhanced Location-Based Services},
JOURNAL = {IEEE Sensors},
YEAR = {2025},
}

@article{dataset,
author = {Sibren De Bast and Sofie Pollin},
journal = {IEEE Dataport},
title = {Ultra Dense Indoor {MaMIMO} {CSI} Dataset},
year = {2021} }

@inproceedings{vaswani2017attention,
  author = {Vaswani, Ashish and others},
  booktitle = {NeurIPS},
  title = {Attention is all you need},
  year = 2017
}

@article{de_bast_expert-knowledge-based_2022,
	title = {Expert-Knowledge-Based Data-Driven Approach for Distributed Localization in Cell-Free Massive {MIMO} Networks},
	journal = {IEEE Access},
	author = {De Bast, Sibren and Vinogradov, Evgenii and Pollin, Sofie},
	year = {2022},
}

@article{tian_attention-aided_2024,
	title = {Attention-Aided Outdoor Localization in Commercial {5G} {NR} Systems},
	journal = {IEEE TMLCN},
	author = {Tian, Guoda and others},
	year = {2024},
}

@inproceedings{tian_high-precision_2023,
	title = {High-Precision Machine-Learning Based Indoor Localization with Massive {MIMO} System},
	booktitle = {IEEE ICC},
	author = {Tian, Guoda and others},
	year = {2023},
}

@article{tian_deep-learning-based_2024,
	title = {Deep-Learning-Based High-Precision Localization With Massive {MIMO}},
	journal = {IEEE TMLCN},
	author = {Tian, Guoda and others},
	year = {2024},
}

@article{dai_deepaoanet_2021,
	title = {{DeepAoANet}: Learning Angle of Arrival from Software Defined Radios with Deep Neural Networks},
	author = {Dai, Zhuangzhuang and others},
	year = {2021},
    journal = {IEEE Access}
}

@inproceedings{tang_regression_2019,
	title = {Regression and Classification for Direction-of-Arrival Estimation with Convolutional Recurrent Neural Networks},
	author = {Tang, Zhenyu and others},
	year = {2019},
    booktitle = {Interspeech}
}

@inproceedings{bast_csi-based_2019,
	title = {{CSI}-based Positioning in Massive {MIMO} systems using Convolutional Neural Networks},
	author = {Bast, Sibren De and Guevara, Adrea P. and Pollin, Sofie},
	year = {2020},
    booktitle = {IEEE VTC-Spring}
}

@inproceedings{bast_mamimo_2020,
	title = {{MaMIMO} {CSI}-based positioning using {CNNs}: Peeking inside the black box},
	author = {Bast, Sibren De and Pollin, Sofie},
	year = {2020},
    booktitle = {IEEE ICC Workshops}
}

@inproceedings{mukherjee_multi-domain_2025,
	title = {Multi-domain {CSI}-based Indoor Localization with Deep Attention Networks for {MIMO} {JCAS} system},
	booktitle = {IEEE JC\&S},
	author = {Mukherjee, Anirban and others},
	year = {2025},
}

@article{kerdjidj_uncovering_2024,
	title = {Uncovering the Potential of Indoor Localization: Role of Deep and Transfer Learning},
	journal = {IEEE Access},
	author = {Kerdjidj, Oussama and others},
	year = {2024},
}

@article{li_toward_2022,
	title = {Toward Fine-Grained Indoor Localization Based on Massive {MIMO}-{OFDM} System: Experiment and Analysis},
	journal = {IEEE Sensors},
	author = {Li, Chenglong and others},
	year = {2022},
}

@article{xu_swin-loc_2024,
	title = {Swin-{Loc}: Transformer-Based {CSI} Fingerprinting Indoor Localization With {MIMO} {ISAC} System},
	journal = {IEEE Trans. Veh. Technol.},
	author = {Xu, Xiaodong and others},
	year = {2024},
}

@inproceedings{ma_millimeter_2024,
	title = {Millimeter Accuracy Indoor Localization System Using an Attention Convolution Model},
	booktitle = {IEEE WCNC},
	author = {Ma, Jiteng and others},
	year = {2024},
}

@inproceedings{hejazi_dyloc_2021,
	title = {{DyLoc}: Dynamic Localization for Massive {MIMO} Using Predictive Recurrent Neural Networks},
	booktitle = {IEEE INFOCOM},
	author = {Hejazi, Farzam and Vuckovic, Katarina and Rahnavard, Nazanin},
	year = {2021},
}

@inproceedings{nguyen_efficient_2023,
	title = {Efficient Spatial-Temporal Angle-Delay Analysis Scheme for Massive {MIMO} Indoor Tracking},
	booktitle = {IEEE ICC},
	author = {Nguyen, Van-Linh and others},
	year = {2023},
}

@inproceedings{nguyen_deep_2023,
	title = {Deep Learning-Based Localization and Outlier Removal Integration Model for Indoor Surveillance},
	booktitle = {IEEE ICC},
	author = {Nguyen, Van-Linh and others},
	year = {2023},
}

\end{document}